\documentclass{kranthi}
\usepackage{graphicx}
\usepackage{amsmath,amsfonts,bm,amssymb}
\usepackage{color}

\url{}

\footlineauthor{Vaikuntanathan, et al.}
\volume{}
\issuenumber{}
\issuedate{}

\begin{document}
\title{Rough interfaces, accurate predictions: The necessity of
  capillary modes in a minimal model of nanoscale hydrophobic
  solvation}

\author{
Suriyanarayanan Vaikuntanathan \affil{1}{Department of Chemistry, The University of Chicago, Chicago, IL, 60637}, 
Grant M Rotskoff \affil{2}{Biophysics Graduate Group, University of California, Berkeley, CA 94720}, 
Alexander Hudson\affil{3}{Department of Chemistry, University of California, Berkeley, CA 94720},
and 
Phillip Geissler\affil{3}{Department of Chemistry, University of California, Berkeley, CA 94720.}
}
\contributor{Submitted to Proceedings of the National Academy of Sciences
of the United States of America}

\significancetext{ 
Hydrophobic effects, which play a crucial role in many chemical and
biological processes, originate in the statistics of microscopic
density fluctuations in liquid water.  Chandler and coworkers have
established the foundation for a simple and unified understanding of
these effects, by identifying essential aspects of water's
intermolecular structure while accounting for its proximity to phase
coexistence.  
Here we show that
coarse-grained models based on this perspective, when constructed to
respect the statistics of capillary waves at interfaces, can achieve
remarkable agreement with results of atomistically detailed
simulations.
Highly efficient and lacking
adjustable parameters, such models hold promise as powerful tools for
studying multiscale problems in hydrophobic solvation.  }

\maketitle

\begin{article}
  \begin{abstract}{
Modern theories of the hydrophobic effect highlight its dependence on
length scale, emphasizing the importance of interfaces in the vicinity
of sizable hydrophobes. We recently showed that a faithful treatment
of such nanoscale interfaces requires careful attention to the
statistics of capillary waves, with significant quantitative
implications for the calculation of solvation thermodynamics. Here we
show that a coarse-grained lattice model like those of Chandler and
coworkers, when informed by this understanding, can capture a broad
range of hydrophobic behaviors with striking accuracy. Specifically,
we calculate probability distributions for microscopic density
fluctuations that agree very well with results of atomistic
simulations, even many standard deviations from the mean, and even for
probe volumes in highly heterogeneous environments. This accuracy is
achieved without adjustment of free parameters, as the model is fully
specified by well-known properties of liquid water. As 
examples of its utility, we compute the free energy profile for a
solute crossing the air-water interface, as well as the thermodynamic
cost of evacuating the space between extended nanoscale surfaces.
These calculations suggest that a highly reduced model for
aqueous solvation can enable efficient multiscale
modeling of spatial organization driven by hydrophobic and interfacial
forces.  }
\end{abstract}


Hydrophobic forces play a crucial role in biological self-assembly, protein folding, ion channel gating, and lipid membrane dynamics~\cite{Chandler2005,Patel2012,Patel2011,Vaikuntanathan2013,Otten2012,Setny2013,Cavallaro2011,Buldyrev2007,Ma2015}. The origin and strength of these forces are well understood at extreme length scales, based on the recognition that accommodating an ideal volume-excluding hydrophobe in water carries the same thermodynamic cost as evacuating solvent from the corresponding volume.  On the scale of a small molecule like methane, density fluctuations that enable such evacuation are Gaussian distributed to a very good approximation, even far from the mean~\cite{Hummer1996}. Linear response theories, such as Pratt-Chandler theory, can thus be quite accurate for assessing solvation of individual small hydrophobic species.

Solvation at much larger scales is by contrast dominated by water's
proximity to liquid-vapor coexistence. Hydrophobic forces involving
extended substrates are shaped by the physics of interfaces, and are
quantified by macroscopic parameters like surface tension.  In
between these extremes, a rich variety of hydrophobic effects results
from the combined importance of nearby phase coexistence and details
of intermolecular structure.  Capturing this interplay, for instance
near a biological macromolecule, presents a significant challenge for
theory.

Lum-Chandler-Weeks (LCW) theory represents the modern understanding of
hydrophobic solvation, providing a conceptual and mathematical
framework to couple interfacial forces with the short-wavelength
density fluctuations that determine solvation of small molecules
~\cite{Hummer1996,Chandler1993}. Its physical perspective has inspired
the development of coarse-grained lattice models, whose applications
have revealed interesting and general mechanisms for the role of water
in hydrophobic self-assembly
processes~\cite{Patel2012,ReintenWolde2001,Varilly2011}.  Primitive
versions of these models, however, long appeared unable to achieve
close quantitative agreement with atomistically detailed simulations,
e.g., for the probabilities of extreme number density fluctuations in
nanometer-scale probe volumes.  This shortcoming motivated the
construction of more elaborate models, which include interactions
between nonadjacent lattice sites~\cite{Varilly2011} and/or explicit
coupling between density fluctuations at short and long
wavelengths~\cite{ReintenWolde2001}. Quantitative accuracy improved as
a result of these additions, but close agreement with detailed
simulations remained elusive, despite the expanded set of adjustable
parameters.

The prominence of interfacial physics in this understanding of
hydrophobic effects suggests that the quantitative success of a
coarse-grained model hinges on its ability to accurately capture the
natural shape fluctuations of a liquid-vapor interface~\cite{Chandler2005,Mittal2008,Aarts2004}. We recently
showed that doing so with LCW-inspired lattice models requires closer
attention to the statistical mechanics of capillary waves than was
previously paid~\cite{Vaikuntanathan2014,Mittal2008}. These fluctuations are
pronounced in molecular simulations, but present in lattice models
only for sufficiently weak coupling $\epsilon$ between lattice sites,
i.e., only at temperatures above the roughening
transition $T_{\rm R}$~\cite{Weeks1973}. In this rough regime the relationship between the
microscopic cohesive energy $\epsilon$ and the macroscopic surface
tension $\gamma$ is nontrivial. This previously unrecognized
connection, which is essential for faithfully representing the
spectrum of capillary waves, yields a lattice model parameterization
that is substantially different than in previous
work. Ref.\@~\cite{Vaikuntanathan2014} thus presented the first
primitive LCW-inspired lattice model that fully respects the
statistical mechanics of capillary waves.

Here, we put the lattice model of Ref.\@~\cite{Vaikuntanathan2014} to a
number of exacting tests, which probe its ability to accurately
describe density fluctuations
on the nanometer scales relevant to protein biophysics. Despite its
coarseness and lack of adjustable parameters, this model
achieves remarkably close agreement with atomistic simulations, even
in scenarios with strong spatial heterogeneity. These tests 
assess the importance of details 
we have omitted,
such as explicit coupling between short- and long-wavelength
density fields. 

A careful 
analysis
of our results underscores
the 
interplay
of length scales accomplished by the
coarse-grained model, highlights the importance of capillary
fluctuations, and emphasizes the special environment for solvation
presented by extended interfaces.
Our ultimate conclusion is that 
diverse
hydrophobic
phenomena can be captured quantitatively at a coarse-grained level,
with minimal attention to atomic-scale intermolecular structure. 
It is sufficient to capture the correct physics at extreme length
scales and link them with simple excluded volume constraints.
We illustrate the promise of such models as practical
tools with an application to the association of hydrophobic plates.

\begin{figure*}[tbp]
\begin{center}
\includegraphics[width=\textwidth]{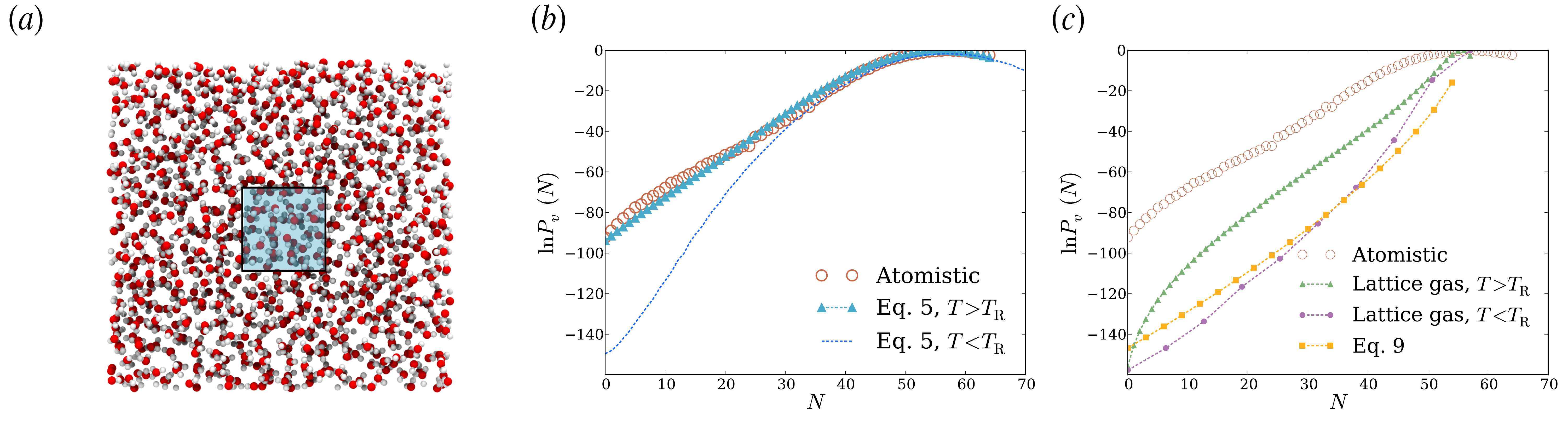}
\end{center}
\label{fig:bulk}
\caption{ Density fluctuations within a cubic probe volume, of size
  12\AA\ $\times$ 12\AA\ $\times$ 12\AA, in bulk liquid water.  (a)
  Cross-sectional snapshot from an atomistic molecular dynamics
  simulation, showing the probe volume $v$ in blue.  (b) and (c)
  Probability distribution $P_v(N)$ of the number of water molecules
  whose center lies in $v$.  Results are shown for atomistic
  simulations, for the LCW-inspired coarse-grained model of
  Eq.\@~\eqref{eq:LCW}, for the conventional lattice gas of
  Eq.\@~\eqref{eq:Ising}, and for the continuum estimate of
  Eq.\@~\eqref{eq:low-density-slope}. Lattice models were simulated with
  two parameter sets consistent with the macroscopic surface tension
  of the air-water interface. The lower temperature case ($T< T_{\rm R}$
  parameters, $\epsilon/T = 6$, $\delta = 4$\AA) does not support
  capillary fluctuations.  The $T>T_{\rm R}$ parameters ($\epsilon/T =
  1.35$, $\delta = 1.84$\AA) by contrast describe an interface whose
  shape fluctuations are consistent with capillary wave
  theory. Results for Eq.\@~\eqref{eq:LCW} with the $T<T_{\rm R}$ parameters,
  plotted in (b), are reproduced from Ref.\@~\cite{Varilly2011}. 
}
\end{figure*}

\section{Coarse-Graining Water} 

Coarse-grained models motivated by the Lum-Chandler-Weeks approach
separately account for density fluctuations at small and large length
scales. As described by Chandler and coworkers in
Refs.\@~\cite{ReintenWolde2001,Varilly2011},
long-wavelength variations are represented on a lattice with
microscopic resolution on the order of a molecular diameter. In the
absence of solutes, walls, or constraints, the
corresponding binary occupation variables $n_i$, which indicate either
vapor-like ($n_i=0$) or liquid-like ($n_i=1$) density in cell $i$, are
governed by a lattice gas Hamiltonian
\begin{equation}
\label{eq:Ising}
H= -\epsilon
\sum_{\langle i,j\rangle} n_i n_j -\mu \sum_i n_i\ , 
\end{equation}
where $\mu$ is the chemical potential for cell occupation, and
$\sum_{\langle i,j\rangle}$ indicates
summation over all pairs of nearest neighbor cells.  We consider
ambient thermodynamic conditions, at which liquid is slightly
more stable than vapor,
$\mu \approx -3\epsilon +\delta \mu$, where $\delta \mu$
is 
the chemical potential offset from coexistence~\cite{ReintenWolde2001}
unless otherwise specified.  As in
Ref.\@~\cite{Vaikuntanathan2014}, we set $\epsilon/T=1.35$, within the
narrow range that is consistent with the statistical mechanics of
rough interfaces yet far from criticality.
The lattice spacing $\delta=1.84$\AA\ is chosen to reproduce the
experimentally determined surface tension $\gamma_w=72 {\rm
  mN/nm^{2}}$ of the air-water interface, according to the approximate
relation
\begin{equation}
\label{eq:deltadef}
\pi \beta \gamma_w \delta^2 = (\beta \epsilon)^2
\end{equation} 
derived in Ref.\@~\cite{Vaikuntanathan2014}. In addition to having the correct surface tension, the power spectrum of interfacial height fluctuations of the lattice gas 
exhibits capillary scaling for this parameter set of $\epsilon,\delta$. 

Regions that are locally liquid-like ($n_i=1$) additionally support
short-wavelength density fluctuations $\delta \rho({\bf r})$, which
are assumed to obey Gaussian statistics~\cite{Chandler1993}.  In the
absence of constraints, these continuous fluctuations are
characterized by the two-point correlation function
\begin{equation}
\label{eq:chidef}
\chi({\bf r}-{\bf r}^\prime)\equiv
\langle \delta\rho({\bf r}) \delta\rho({\bf r}')\rangle 
=
\rho_\ell \delta({\bf r}-{\bf
  r}^\prime)+\rho_\ell^2(g({\bf r}-{\bf r}^\prime)-1)\ ,
\end{equation}
where ${\bf r}$ and ${\bf r}'$ label positions inside the liquid,
$\rho_{\ell}$ is the macroscopic number density of pure liquid water,
and $g(r)$ is the radial distribution function~\cite{Rowlison2002}.
We use estimates of $g(r)$ and its Fourier transform obtained from experimental measurements by Narten and Levy~\cite{Narten1971}.

We consider ideal hydrophobic solutes, whose influence on
the solvent is to exclude it from a volume $v$.  Weak, smoothly
varying attractive interactions between solute and solvent amount to a
small perturbation in this context. The effect of, e.g., dispersion
forces can therefore be reasonably addressed using perturbation
theory~\cite{ReintenWolde2001,Varilly2011}. 

We model such idealized solutes by imposing a constraint of solvent
evacuation: The total density within
$v$ must vanish, 
\begin{equation}
\sum_{i\in v} \left[n_i \rho_\ell v_i +\int_{v_i} d{\bf r}\ \delta
    \rho({\bf r})\right]=0,
\label{eq:densitydef}
\end{equation} 
where the sum runs over all lattice cells $i$ that intersect $v$,
and $v_i$ is the corresponding volume of intersection.

Gaussian fluctuations in the rapidly varying field $\delta\rho({\bf
  r})$ can be integrated out exactly~\cite{Chandler1993}, yielding an
effective Hamiltonian $H_v[\{n_i\}]$ 
for the lattice occupation variables.  In the
presence of $m$ ideal volume-excluding
solutes~\cite{ReintenWolde2001,Varilly2011},
\begin{equation}
\label{eq:LCW}
H_v[\{n_i\}]= -\epsilon\sum_{\langle i,j\rangle} n_i n_j - \mu \sum n_i +  
       \frac{T}{2} \left[{\overline{N}^{\top}} {\chi^{-1}_{\rm in}}  {\overline{N}} +C\right]\,, 
\end{equation}
where $\overline{N}$ is an $m$-component
column vector with elements
\begin{equation}
{N_{\alpha}}=\Sigma_i \rho_l n_i v^{(\alpha)}_i\,.
\label{eq:LCWdetails2}
\end{equation}
$v^{(\alpha)}_i$ 
denotes the volume of overlap between the $\alpha^{\rm th}$ solute
and lattice cell $i$, and ${\chi}_{\rm in}$ is an $m\times m$ square
matrix with elements
\begin{equation}
\left({\chi}_{\rm in}\right)_{\alpha,\beta}=\int_{{\bf r}\in v^{(\alpha)}}\int_{ {\bf r}^\prime\in v^{(\beta)}}  \Theta({\bf r}) \chi({\bf r},{\bf r}^\prime) \Theta({\bf r}^\prime)\,. 
\end{equation}
Here, $\Theta ({\bf r})=1$ if the lattice cell containing ${\bf r}$ is
occupied by solvent and vanishes otherwise. The constant $C$ is given by 
\begin{equation}
  C=\begin{cases}
    \ln\left(\det\left(2\pi\chi_{\rm in}\right)\right) & \text{if $ \sum_{\alpha} \overline{N}_{\alpha} > 1$},\\
    \max \left[\ln\left(\det\left(2\pi\chi_{\rm in}\right)\right), \sum_{\alpha} \overline{N}_{\alpha} \right] & \text{otherwise}\,.
  \end{cases}
\end{equation}

The last two terms in Eq.\@~\eqref{eq:LCW} improve upon previous
lattice-based models, providing a computationally tractable yet
quantitatively accurate approximation for solute-solute interactions
mediated by Gaussian density fluctuations in the surrounding solvent.
Starting from a Gaussian field theory~\cite{Chandler1993}, they
may be derived by applying the
constraint in Eq.\@~\eqref{eq:densitydef} separately to each solute's
excluded volume. Details of this derivation are included in the
Supporting Information.

Note that the coarse-grained model defined by Eq.\@~\eqref{eq:LCW} 
has
essentially no free parameters---in light of Eq.\@~\eqref{eq:deltadef}
there is very little freedom in
the choice of $\epsilon$ and $\delta$~\cite{Vaikuntanathan2014}.
We only require the surface tension of water
and the bulk radial distribution function
$g(r)$. 
The lattice
model also lacks explicit energetic coupling between the density
fields $n_i$ and $\delta \rho({\bf r})$. Their interdependence arises
strictly from the excluded volume constraint expressed in
Eq.\@~\eqref{eq:densitydef}.

Below we compare calculations based on this coarse-grained model with
results of atomistic molecular dynamics simulations. Molecular
simulations were performed using the SPC/E model of water
(Methods). Several of these calculations scrutinize the extreme wings
of probability distributions, which were accessed using standard
techniques of umbrella sampling (Methods). 

\section{Results and discussion}
We have performed calculations that stringently assess the ability of
an LCW-inspired lattice model to capture details of hydrophobic
solvation at small, large, and intermediate length scales.  We focus
on characterizing and comparing the statistics of density fluctuations
within microscopic probe volumes, in part because of their direct
relevance to solubility. Specifically, we calculate the probability
$P_v(N)$ of observing $N$ solvent molecules within a probe volume of
size $v$. Its extreme value determines the excess chemical potential
$\mu_{\rm ex}(v)$ of an ideal hydrophobe with the corresponding
excluded volume, $\mu_{\rm ex}(v) = -T \ln P_v(0)$. The behavior of
$P_v(N)$ between this extreme case ($N=0$) and more typical values ($N
\approx \rho_\ell v$) reveals much about the physical nature of
fluctuations that might be accessed through application of external
fields, solute attractions, or changes in thermodynamic state~\cite{Patel2012,Patel2011}. With 
these implications in mind, we have computed $P_v(N)$ over its
entire meaningful range for a variety of scenarios pertinent to
solvation in complex environments.

In addition to atomistically detailed molecular dynamics simulations
and our LCW-inspired lattice model, we present results for several
less sophisticated models. These simpler descriptions lack one or more
of the physical ingredients underlying LCW theory, and thus shed light
on their relative importance.  For example, short-wavelength
fluctuations can be straightforwardly neglected by studying the
conventional lattice gas described by Eq.\@~\eqref{eq:Ising}, which lacks
biases from Gaussian fluctuations in $\delta\rho({\bf r})$. In this
case density variations within a probe volume can be achieved only
through fluctuations of the binary occupation variables $n_i$. For the
parameterization we have described ($\epsilon/T = 1.35$, $\delta =
1.84$\AA), which ensures $T>T_{\rm R}$, interfaces of this
lattice gas
exhibit a spectrum of capillary modes comparable to that of
a natural liquid-vapor interface. Fluctuations and response of these
capillary modes may contribute significantly to solvation structure
and thermodynamics~\cite{Vaikuntanathan2013,Otten2012,Setny2013}.

To assess the importance of capillary fluctuations, we examine a
different parameterization of Eq.\@~\eqref{eq:Ising} ($\epsilon/T = 6$,
$\delta = 4$\AA), for which $T<T_{\rm R}$.  This lattice
gas too supports interfaces with the correct surface tension.
Because its roughening transition lies above
ambient temperature, however, the lattice gas with $T<T_{\rm R}$ 
lacks fluctuations
in surface topography characteristic of capillary modes (i.e., the
typical amplitude of long-wavelength undulations is not proportional
to their wavelength).  Results from this model thus isolate the
contribution of interfacial flexibility to hydrophobic effects, a
flexibility that is also neglected in the mean field treatment of LCW
theory~\cite{Vaikuntanathan2014}.

Together, these calculations explore the 
interplay between short- and long-wavelength aspects of
hydrophobicity. We find in general that a simple LCW-inspired lattice
model can describe with surprising accuracy the statistics of density
fluctuations observed in detailed molecular simulations. This success
is compromised substantially in most cases by omitting the effects of
short-wavelength fluctuations and/or capillary waves, suggesting that
our coarse-grained model contains a minimum of microscopic detail
required to quantitatively capture the solvation and association of
nanoscale hydrophobic species.

\subsection{Statistics of density fluctuations in bulk water}
We first examine density fluctuations in the simplest aqueous
environment, i.e., bulk liquid water. It has been well established by
MD simulations that for small probe volumes ($v \lesssim 0.5$~nm$^3$)
in this homogeneous setting, $P_v(N)$ has a Gaussian form well into
its tails~\cite{Hummer1996}. For larger $v$, low-density fluctuations
are strongly biased by the small chemical potential difference between
macroscopic liquid and vapor.
$P_v(N)$ then develops an
exponential tail, decaying much more slowly than 
the Gaussian
fluctuations near $\langle N \rangle = \rho_\ell v$ would suggest~\cite{Varilly2011}.
These basic features of $P_v(N)$ are essentially built into
LCW-inspired models, but their details can be quite sensitive to the
way such models are constructed and parameterized. As shown in
Fig.\@~\ref{fig:bulk}(b), the model of Eq.\@~\eqref{eq:LCW}, when
parameterized with attention to capillary fluctuations, does an
excellent job reproducing distributions obtained from atomistic MD
simulations for nanometer-scale
cubic probe volumes, over a very wide range of $N$.

Reversibly decreasing $N$ in atomistic simulations from its average
value induces formation of a small, roughly cubic cavity that grows to
span $v$ as $N\rightarrow 0$. This scenario suggests a simple
continuum estimate of $P_v(N)$ that resolves only the growing
interfacial area of the cavity as the probe volume is evacuated. The
surface area of a cubic cavity which accommodates an average of $N$
water molecules in bulk is $A = 6 (N/\rho_l)^{2/3},$
which we use as an estimate of the surface area of the cavity that 
appears as water molecules evacuate the probe volume.
Assigning the macroscopic surface tension $\gamma_w$ as
the free energy cost per unit area of the microscopic cavity, we
obtain a prediction for the decay rate of the exponential tail of
$P_v(N)$,
\begin{equation}
\label{eq:low-density-slope}
\frac{\partial \ln P_v(N)}{\partial N} \approx 4 \beta \gamma_w \left(\frac{\rho_l^2}{\langle N \rangle -N}\right)^{1/3},
\end{equation} 
that agrees reasonably well
with detailed simulation results. Lacking sensitivity to microscopic
fluctuations, this estimate (plotted in Fig.\@~\ref{fig:bulk}(c)) unsurprisingly fails to capture the
Gaussian character of $P_v(N)$ near $\langle N \rangle$. Nor does it
describe well the overall free energy scale associated with emptying
$v$, erring by more than 70 $k_{\rm B}T$.

Fig.\@~\ref{fig:bulk}(c) also shows results obtained from simulations
of the lattice gas with $T<T_{\rm R}$. Its prediction for $P_v(N)$ closely resembles
the simple continuum estimate of Eq.\@~\eqref{eq:low-density-slope},
accurately describing the low-density slope of $\ln P_v(N)$ but not
its scale or peak behavior. This similarity highlights limitations of
lattice models at temperatures below the roughening transition temperature. The
deficiency of spontaneous fluctuations in interfacial shape begets an
overly stiff response to fields or constraints imposed by
solutes. Lattice degrees of freedom serve here only to coarsely
determine static interfaces when solutes are large enough to induce
drying.  Lack of capillary modes further renders the surface tension
of a lattice gas below the roughening transition temperature anisotropic, introducing the possibility of
strong lattice artifacts. In this case of a nanometer-scale cubic
probe volume in bulk liquid, correspondence with the continuum
estimate suggests that such artifacts are not substantial.

The parameterization of Eq.\@~\eqref{eq:Ising} we have advocated,
which does capture capillary fluctuations at the liquid-vapor
interface, significantly improves agreement between atomistic
simulations and the conventional lattice gas.
Plotted in Fig.\@~\ref{fig:bulk}(c), the $T>T_{\rm R}$ lattice gas result for
$P_v(N)$ manifests low-density fluctuations that are dramatically more
probable than for the lattice gas with $T<T_{\rm R}$. Agreement with atomistic
simulations nonetheless remains very poor, signaling a critical role
for short-wavelength modes even in the exponential tail of $P_v(N)$.

The success achieved by the full coarse-grained model of
Eq.\@~\eqref{eq:LCW} is thus not a transparent consequence of the
limiting behaviors motivating its form. Instead, a subtle cooperation
of interfacial fluctuations and thermodynamics, together with Gaussian
density statistics at the molecular scale, underlies its accurate
prediction for $P_v(N)$ across the entire range of $N$.

LCW-inspired models based on the $T< T_{\rm R}$ lattice gas are much more
difficult to reconcile with atomistic simulations.  When parameterized
with $T<T_{\rm R}$, the unadorned form of Eq.\@~\eqref{eq:LCW} accurately
predicts $P_v(N)$ only near its peak, failing dramatically at low $N$
where interfacial fluctuations figure prominently. (See Fig.\@~SI~1.)
Ref.\@~\cite{Varilly2011} outlines two strategies to address this
shortcoming.  Smearing out discrete interfaces with a numerical
interpolation scheme improves predictions substantially, but still
fails to achieve quantitative accuracy in the extreme tail of
$P_v(N)$. Adding as well an estimate of unbalanced attractive forces produces near quantitative agreement~\cite{Varilly2011}. These elaborations,
however, require introducing interaction potentials and adjustable
parameters that are not clearly specified by experimental
measurements~\cite{Varilly2011}.  Our results show that greater
accuracy can be achieved much more simply, by using lattice gas
parameters that properly represent the statistics of capillary
fluctuations.

\subsection{Density fluctuations at a liquid-vapor interface}
The interplay of physical factors determining hydrophobic solvation
can resolve much differently in spatially heterogeneous environments.
To explore basic effects of such nonuniformity, we have examined
microscopic density fluctuations at the interface between air and
water.
Specifically, we
consider a cubic probe volume that straddles the plane of a
macroscopic phase boundary, i.e., the Gibbs dividing surface between
liquid and vapor. The overall shape of $P_v(N)$ in this case is
similar to the bulk result, featuring Gaussian statistics near the
peak and a more slowly decaying low-density tail. In this case,
however, the thermodynamic cost of evacuation is much lower than in
bulk, despite a similar value of $\langle N \rangle$.

\begin{figure}[h]
\begin{center}
\includegraphics[width=\linewidth]{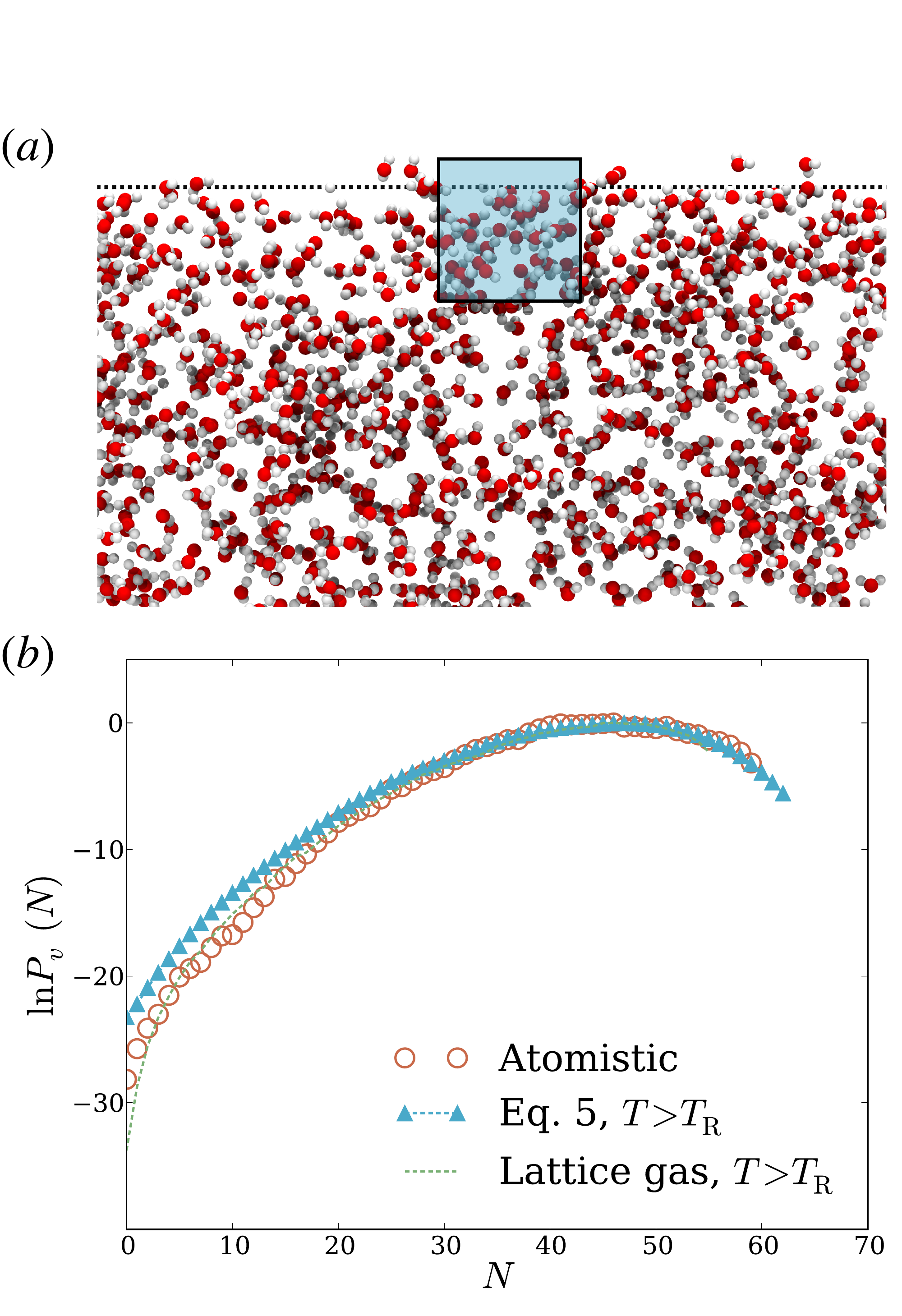}
\end{center}
\caption{ 
Density fluctuations
  within a cubic probe volume that straddles the interface between
  liquid water and its vapor.  The probe volume, of size
  12\AA\ $\times$ 12\AA\ $\times$ 12\AA,
  is centered slightly within the liquid phase, 3.67\AA\ below
  the Gibbs' dividing surface.
  (a) Cross-sectional snapshot from an atomistic molecular dynamics simulation, 
  showing the probe volume $v$ in blue.  
  (b) Probability distribution $P_v(N)$ of the number of water molecules
  whose center lies in $v$.  Results are shown for atomistic
  simulations, for the LCW-inspired coarse-grained model of
  Eq.\@~\eqref{eq:LCW} at coexistence, and for the conventional lattice gas of
  Eq.\@~\eqref{eq:Ising} at coexistence. Lattice models were simulated with
  $T>T_{\rm R}$
  parameters ($\epsilon/T = 1.35$, $\delta=1.84$\AA) that yield both the 
  correct surface tension and capillary wave scaling.  }
\label{fig:interface}
\end{figure}

The LCW-inspired lattice model of Eq.\@~\eqref{eq:LCW} again matches
atomistic simulation results very well, both near the mean of $P_v(N)$
and far into its low-density wing.  Our reduced description is
therefore a promising tool for assessing hydrophobic solvation near
the liquid's boundary.

While the shape of $P_v(N)$ we have determined for the interfacial
environment resembles that of bulk liquid, the underlying structural
fluctuations are quite different. This difference is made clear by
considering the simple lattice gas (in its higher-temperature
parameterization), whose prediction is also plotted in
Fig.\@~\ref{fig:interface}(b). In contrast to our bulk liquid results,
neglecting short-wavelength density fluctuations in this case effects
only a modest suppression of extreme low-density excursions; the shape
and scale of $P_v(N)$ are in fact captured well by the lattice
gas with $T>T_{\rm R}$. Correspondingly, a calculation based entirely on short-wavelength
fluctuations, with a static, flat interface, fails to capture the shape
of $P_v(N)$ even near its peak. (See Fig.\@~SI~1.)

The long-wavelength density component 
thus dominates the
response of the LCW-inspired model in this spatially heterogeneous
scenario, highlighting the key importance of capillary fluctuations at
the air-water interface.
Evacuation of
a probe volume can be inexpensively achieved near a pre-existing
interface by simply deforming its shape.  Refs.\@~\cite{Patel2010}
and~\cite{Patel2012} have also pointed to interfacial deformation as a
mechanism for extreme density fluctuations near ideal hydrophobic
surfaces and hydrophobic biological molecules.

The statistics of finer scale density variations have non-negligible
quantitative impact on the predictions of Eq.\@~\eqref{eq:Ising} (e.g.,
reducing the cavitation free energy by roughly 5 $k_{\rm B}T$) 
but do
not qualitatively shape the solvent response as in the bulk
case. Underscoring the role of surface shape fluctuations, the
lower-temperature parameterization of the lattice gas, which lacks
capillary waves, fails profoundly to describe occupation statistics
for the interfacial probe volume, as shown in Fig.\@~SI~1.

\begin{figure}[h]
\begin{center}
\includegraphics[width=\linewidth]{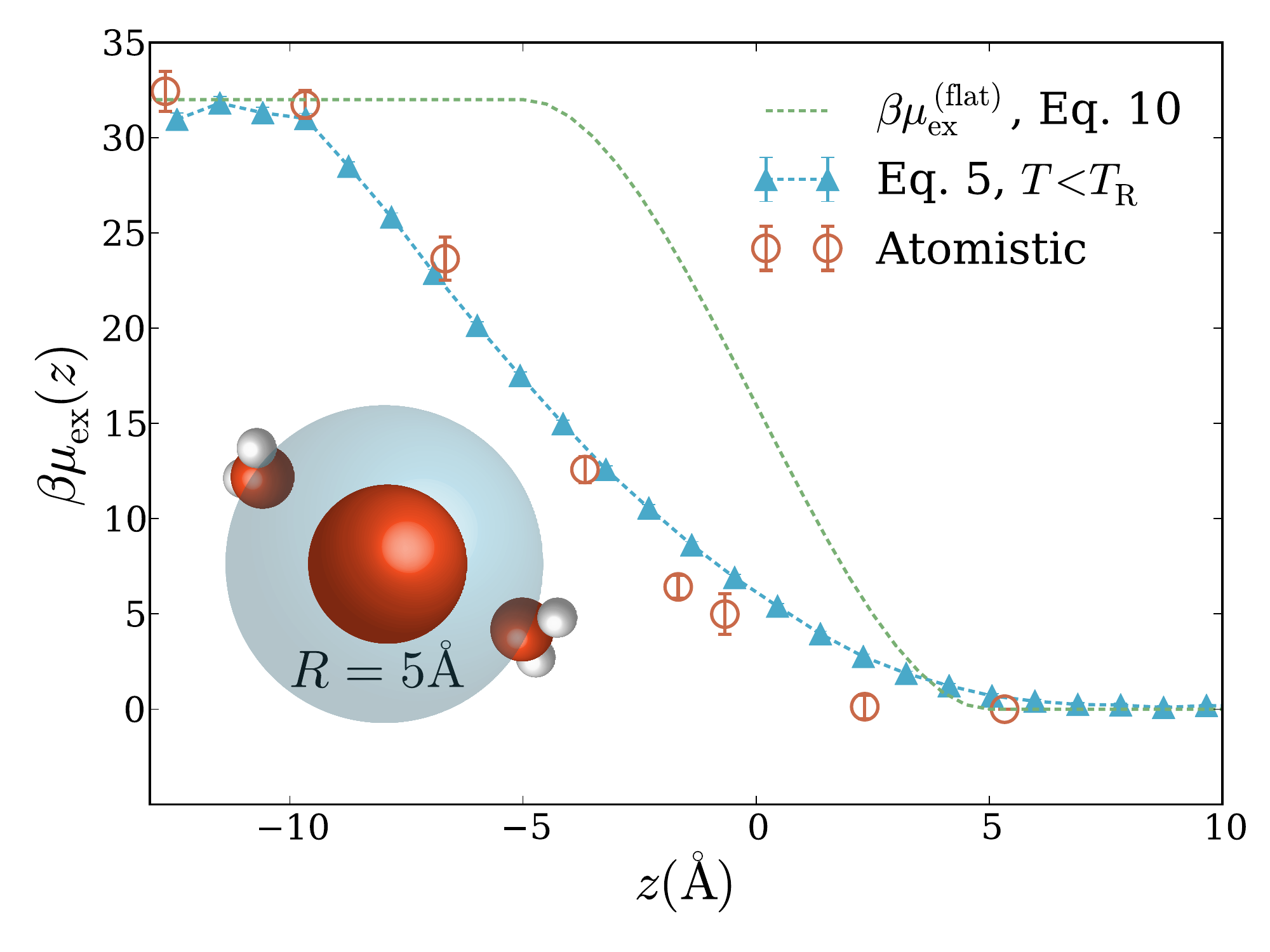}
\end{center}
\caption{Excess chemical potential of a spherical hydrophobic solute
  as a function of its perpendicular displacement $z$ from the
  air-water interface. 
  $z\rightarrow \infty$ corresponds to bulk vapor,
  $z\rightarrow -\infty$ to bulk liquid, and $z=0$ to the Gibbs
  dividing surface between the two coexisting phases.  Data are shown
  for atomistic simulations, for the LCW-inspired coarse-grained model
  of Eq.\@~\eqref{eq:LCW} (with $T>T_{\rm R}$ parameters $\epsilon/T = 1.35$, 
  $\delta=1.84$\AA), and for the estimate in Eq.\@~\eqref{eq:muexflat} based 
  on a completely quiescent interface.  This solute excludes the center of
  each water molecule from a sphere of radius $R=5$\AA\ (as
  shown in the inset schematic).
  }
\label{fig:solvation}
\end{figure}

\begin{figure*}[tbp]
\begin{center}
\includegraphics[width=\textwidth]{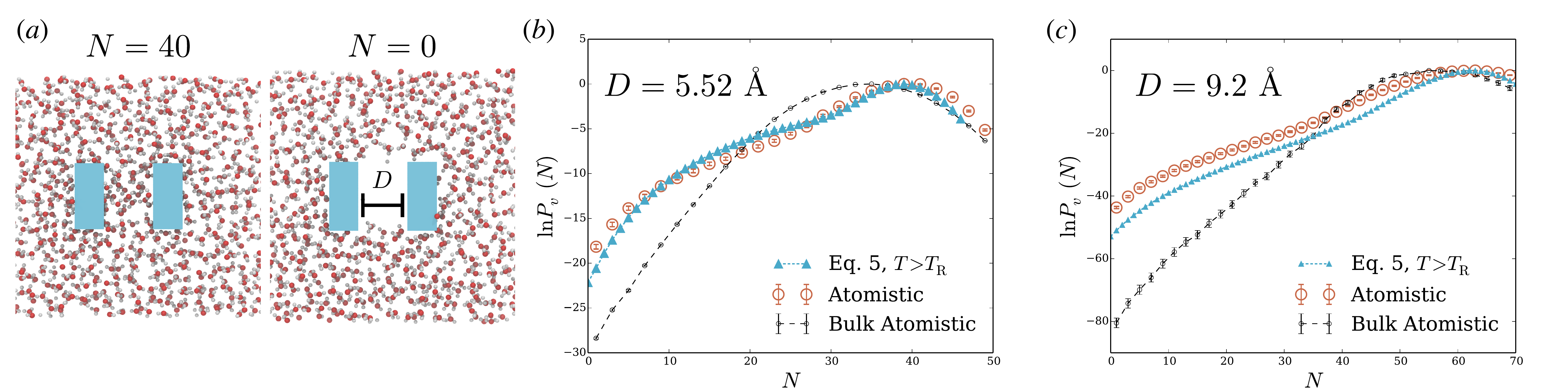}
\end{center}
\caption{ Statistics of aqueous density fluctuations between
  nanometer-scale hydrophobic plates immersed in the liquid phase.
  (a) Cross-sectional snapshot from an atomistic molecular dynamics
  simulation, showing the volume-excluding hydrophobic plates in blue.
  (b) and (c) Probability distributions $P_v(N)$ of the number of
  water molecules whose center lies between the two plates.  Results
  are shown for atomistic simulations and for the LCW-inspired
  coarse-grained model of Eq.\@~\eqref{eq:LCW} (with $T>T_{\rm R}$
  parameters  $\epsilon/T = 1.35$, $\delta=1.84$\AA). 
  We also plot results for a probe
  volume of equivalent size in bulk water, obtained from atomistic
  simulations.  The plates are cuboids that exclude the center of each
  water molecule from a volume with dimensions
  20.24\AA\ $\times$ 5.52\AA\ $\times$ 9.2\AA,
  and are separated along their short dimension
  by a distance $D$. Plate separations were
  fixed at $D = 5.52$\AA\ in the simulations of panel (b), and at $D =
  9.2$\AA\ for (c).  }
\label{fig:plates}
\end{figure*} 

\subsection{Association of a hydrophobic solute with the interface}

The occupation statistics discussed above hint at thermodynamic
driving forces that 
govern
solvation of nonpolar species in
interfacial environments. To make this connection explicit, we have
computed the excess chemical potential $\mu_{\rm ex}(z)$ of a
spherical hydrophobe as a function of its perpendicular displacement
$z$ from the air-water interface (with $z\rightarrow \infty$
indicating bulk vapor and $z\rightarrow -\infty$ bulk liquid). Such
interfacial free energy profiles
have been determined from molecular simulations for a variety of
solutes~\cite{Vaikuntanathan2013,Otten2012,Jungwirth2006,Venkateshwaran2014}, including small ions, which can exhibit a surprising
tendency to adsorb to the liquid-vapor phase boundary~\cite{Otten2012}.  While
charged solutes are distinct in important ways from hydrophobes, the
cost of creating solute-sized cavities has been implicated as a key
factor in their surface affinity~\cite{Vaikuntanathan2013,Otten2012}. As a simple estimate of this
cost, Levin et al.\ have assigned a fixed solvation free energy $f_{\rm
  disp}$ per unit of solvent volume displaced by the
  solute~\cite{Levin2009}. For an ideally flat interface, this approximation yields a
free energy profile,
\begin{equation}
\label{eq:muexflat}
\mu_{\rm ex}^{(\rm flat)}(z) 
= f_{\rm disp}\left[\frac{1}{2}-\frac{ z\left(3 R^2-z^2\right)}{4 R^3}\right]\,\,\,,-R\leq z\leq R,
\end{equation}
that is antisymmetric about $z=0$ (once $\mu_{\rm ex}(z)$ has been
shifted by its value at $z=0$).  This estimate asserts that the
interface influences solubility only for distances $|z|$ smaller than
the solute radius $R$. It clearly neglects the role of capillary
fluctuations, which we have argued can be the dominant mode of solvent
response in such scenarios.

Atomistic simulations show that Eq.\@~\eqref{eq:muexflat} poorly approximates the free
energy profile for an ideally hydrophobic nanometer-scale solute.  As
shown in Fig.\@~\ref{fig:solvation} for $R = 5$\AA, the atomistic potential of mean force
is highly asymmetric about $z=0$. On the liquid side ($z<0$), the
interface's influence extends in distance $z$ well beyond the solute's
radius, with $\mu_{\rm ex}(z)$ deviating appreciably from $\mu_{\rm
  ex}(-\infty)$ out to nearly $z=-2 R$. This extended range strongly
implicates variations in surface topography away from the
microscopically flat geometry assumed by Eq.\@~\eqref{eq:muexflat}.

Our LCW-inspired lattice model, by contrast, faithfully captures these
features of $\mu_{\rm ex}(z)$. (See Fig.\@~\ref{fig:solvation}.) Its
minimal ingredients thus appear sufficient to accurately describe the
thermodynamics of accommodating volume-excluding solutes near the
liquid's boundary.  Given the dominant role of interfacial softness in
determining $P_v(N)$ for the lattice model, 
an
accounting of capillary fluctuations appears essential for assessing the
surface affinity of nonpolar solutes. At the same time, these results
suggest that a volume-excluding hydrophobe may be a problematic
reference system for understanding interfacial solvation of charged
species: As a cavity near the interface is charged, strong response
will be induced not only in solvent polarization, as accounted by
dielectric continuum theory in Ref.\@~\cite{Levin2009}, but also in interfacial shape,
an aspect not addressed in existing theories for interfacial ion
solvation.

\subsection{Density fluctuations between ideal hydrophobic plates}
The calculations described so far demonstrate that a simple numerical
implementation of the LCW perspective can accurately describe
hydrophobic effects involving both microscopic structural response and
macroscopic bistability, featuring interfaces that may be pre-existing
or emergent. This success encourages use of such a model to address
more complicated and specific situations that arise in modern
biophysics and materials science, e.g., 
water flow in nanotubes~\cite{Hummer2001}, 
gating of transmembrane ion channels~\cite{Zhu2010}, or the development
 of tertiary and quaternary protein structure~\cite{Berne2009}. 
 In each of these
phenomena, large-scale atomistic simulations have revealed intriguing
functional roles for hydrophobic response as solute configurations
rearrange.  Towards these frontier applications, we consider as a
final example aqueous density fluctuations in a confined hydrophobic
environment.

When confined at the nanometer scale or below, water can exhibit
physical properties markedly distinct from those of the bulk liquid~\cite{Rasaiah2008}.
Interactions with the containing boundaries become a
critical consideration, with hydrophilic and hydrophobic walls
generating very different structural motifs and susceptibilities.
Hydrophobic walls are known to greatly enhance density fluctuations,
so that even weak external fields can induce nanoscale drying~\cite{Patel2011}.  By
poising water near a highly cooperative transition, such constraints
can thus be used to engineer switching under minor perturbation, which
in turn can sharply modulate functional behaviors like transport and
self-assembly~\cite{Chandler2005}.

Using atomistic and coarse-grained simulations, we examined a model
confinement scenario featuring two ideally hydrophobic parallel plates
($11 \delta \times 3 \delta \times 5 \delta$ in size), separated along
their short dimension by a distance $D$, immersed in liquid water, as
depicted in Fig.\@~\ref{fig:plates}(a). As in our other examples, we
focus on occupation statistics of a probe volume, here comprising the
space between the two plates (a volume of size $11\delta \times
5\delta \times D$). Results for $P_v(N)$ are plotted in
Fig.\@~\ref{fig:plates} for two separations, $D=3\delta=5.52$\AA\ (b)
and $D=5\delta=9.2$\AA\ (c). For comparison we also show the
corresponding probability distributions for a probe volume of the same
size and shape placed in homogeneous bulk water, obtained from
atomistic simulations.

The presence of these hydrophobic plates is not sufficient to induce
drying for either value of $D$ considered. The average density between
plates is in fact greater than in bulk water, due to the tendency of
molecules in dense liquids to pack tightly against hard walls.  This
modest elevation of $\langle N \rangle$ is recapitulated by the
LCW-inspired lattice model.  Such an increase in local density
is also captured by a purely Gaussian model that considers only
short-wavelength density fluctuations, underscoring its origin in
simple packing effects. The simple Gaussian model, however,
considerably overestimates the magnitude of this shift, shown in
Figs.\@~SI~2 and~SI~3---while these plates are not sufficiently
confining to evacuate the probe volume with high probability, they do
significantly enhance lattice fluctuations in their vicinity.

The hydrophobic plates have a much stronger impact on the tails of
$P_v(N)$. For the smaller separation, $D=3\delta$, extreme low-density
fluctuations are substantially more probable than in the bulk case,
reflecting stabilization of the dry state.  Typical configurations do
not manifest this stabilization; it would instead be apparent in the
response to an external field that disfavors occupation of the
inter-plate region. Here, a field strength of just 0.5 $T$ per 
molecule
would be sufficient to induce drying.  This profound impact of
hydrophobic confinement on susceptibility to weak perturbations,
despite negligible influence on typical fluctuations, has been
emphasized in previous work, and explored in detail in the context of
protein complex formation~\cite{Patel2012,Berne2009,Rasaiah2008}.

These behaviors too are well described by the LCW-inspired lattice
model.  The slope of $\ln P_v(N)$ in the low-density range, which
largely determines the susceptibilities discussed above, is predicted
especially accurately.  The emergence of an inflection point, a sign
of incipient bistability as the plates approach, is also accurately
captured by the coarse-grained model.
The difficulty of describing solvent-mediated attraction between hydrophobes using linear response theory has been detailed by others~\cite{Pratt2002,Chaudhari2013}. 
For the nanoscale plates considered here, the addition of fluctuating lattice degrees of freedom achieves this goal with striking success.

\section{Conclusions}

Hydrophobic effects drive the formation of diverse 
assemblies in biological and materials systems.
Our results suggest
that the microscopic basis of these effects is thoroughly described by
the physical perspective put forth by Lum, Chandler, and
Weeks. 
Statistics of extreme fluctuations that determine solvation
thermodynamics can be captured with quantitative accuracy in a
lattice model based on the LCW perspective. Doing so, however,
requires careful attention to the softness of air-water interfaces, a
property lacking in many previous models.

The lattice model defined by Eq.\@~\eqref{eq:LCW} 
appears to be truly minimal
for this purpose. Omitting any of its contributions degrades the close
agreement with atomistic simulations we have demonstrated. Moreover,
its parameters are highly constrained by basic experimental
observations, namely surface tension, molecular pair correlations, and
the spectrum of long-wavelength capillary waves.

Notably absent in our model are several ingredients introduced in
previous studies to improve agreement with detailed molecular
simulations.  
We have not introduced explicit coupling between the rapidly and
slowly varying components of the density field, i.e., between the
lattice variables $n_i$ and the Gaussian field $\delta \rho({\bf r})$.
Others have motivated such coupling from the form of self-consistent
equations in a mean-field treatment of the slowly varying density
component~\cite{Lum1999}, which manifest unbalanced attractive forces
due to the rapidly varying component.  Our omission does not, however,
imply a lack of unbalanced attraction in the model of
Eq.\@~\eqref{eq:LCW}.  Direct interactions among the lattice variables
$n_i$ in Eq.\@~\eqref{eq:Ising} are of course sufficient to stabilize interfaces, the
primary and essential role of unbalanced forces in the mean field
theories of Refs.\@~\cite{Lum1999} and~\cite{Weeks2002}. Rather,
Eq.\@~\eqref{eq:LCW} neglects the specific source of unbalanced
attraction due to short-wavelength structure, a coupling whose form
and strength are not transparent for an associated liquid like
water~\cite{Weeks2002}.  Nor does our omission imply a lack of
coupling between $n_i$ and $\delta \rho({\bf r})$.  The manipulations
leading to Eq.\@~\eqref{eq:LCW} permit short-wavelength density
fluctuations only in regions that are liquid-like ($n_i=1$). 
By correlating the statistics of $\delta \rho({\bf r})$
with lattice fluctuations that support phase coexistence,
this constraint effects a potent but
implicit interaction across length scales.  Finally, in LCW-inspired
lattice models explicit length scale coupling has the side effect of
altering statistics of Gaussian density fluctuations in the vicinity
of inhomogeneous lattice configurations.  According to molecular
simulations, short-wavelength fluctuations in liquid regions are in
fact quite robust against such heterogeneity, particularly when liquid
domains are identified with sensitivity to interfacial
fluctuations~\cite{Patel2010,Willard2010}.
  
The lattice gas on which our model is built includes only nearest
neighbor interactions. As emphasized in Ref.\@~\cite{Varilly2011}, 
the ground state of this model
possesses unphysical degeneracies in the shape of closed, convex
interfaces. 
These degeneracies can be removed by introducing lattice interactions
between non-neighboring cells. In a lattice gas below its roughening transition temperature the impact of
these additional interactions is dramatic, since they suppress the
only affordable mode of interfacial shape variation. Our studies of
lattice gases above the roughening transition temperature
suggest that such degeneracies are much less important
in the presence of natural capillary fluctuations.

Such additional couplings may be needed to further improve quantitative predictions, or to address more complicated scenarios. 
These goals may also require attention to molecular details that have not yet been incorporated into LCW-inspired models, for instance concerning the geometry of hydrogen bonds, coordination statistics, or the specific form of interaction potentials.
Pratt and coworkers have made significant advances towards understanding and quantifying the role of these effects in hydrophobicity~\cite{Pratt2002}.
 Incorporating them into a lattice-based model poses significant challenges. 

In addition to helping establish a conceptual foundation for complex
hydrophobic phenomena, our studies advance the more pragmatic goal of
faithfully simulating systems that comprise very large numbers of
water molecules. This challenge limits, for example, the scale of
biomolecular problems that can be examined by simulation without
reducing the description of solvent fluctuations to a gross
caricature.  For the systems we have discussed, our coarse-grained
approach reduces the computational cost of representing explicit
solvent fluctuations by more than two orders of magnitude (relative to
atomistic simulations), while preserving microscopic realism with
surprising accuracy.  This advantage should become even more
significant for very large systems, whose computational burden scales
exactly linearly in Eq.\@~\eqref{eq:LCW}.  The model we have described
could thus enable study of problems that currently lie outside the
reach of atomistically detailed simulations.

\begin{acknowledgments}
SV and PLG were supported by the US Department of
Energy, Office of Basic Energy Sciences, through the Chemical Sciences
Division (CSD) of the Lawrence Berkeley National Laboratory (LBNL),
under Contract DE-AC02-05CH11231. SV was supported by the University
of Chicago for later stages of this project. GMR was supported by an
NSF Graduate Research Fellowship. AH was supported by an NIH training
grant under award number T32GM008295. We acknowledge computational
resources obtained under NSF award CHE-1048789. Computing resources
of the Midway-RCC computing cluster at University of Chicago are also
acknowledged.
\end{acknowledgments}

\section*{Methods}
All atomistic simulations included 6,912 rigid water molecules,
interacting through the SPC/E potential~\cite{Berendsen1987}, in a
periodically replicated simulation cell with dimensions
75\AA\ $\times$ 75\AA\ $\times$ 100\AA,
and held at temperature $300$~K using a 
Nos\'{e}-Hoover thermostat~\cite{Nose1984,Hoover1985}. These conditions
enforce phase equilibrium, with coexisting slabs of liquid and vapor
as in Ref.\@~\cite{Patel2010}.  Probe volumes were positioned at least
10\AA\ away from the resulting liquid-vapor interface, except where
indicated otherwise. Electrostatic interactions were summed using the
particle mesh Ewald algorithm~\cite{Darden1993}.  Intramolecular
constraints were imposed using the SHAKE algorithm~\cite{Ryckaert1977}.
Molecular dynamics trajectories were advanced in
time using the LAMMPS software package~\cite{Plimpton1995}. Probability
distributions for occupation statistics of a probe volume were
determined with umbrella sampling using the INDUS technique~\cite{Patel2010}.

Coarse-grained simulations were performed with an in-house software
package that is available upon request. These systems comprised
$30\times 30 \times 30$ lattice cells, periodically replicated in each
direction.
Probability distributions of the lattice occupation state within a
probe volume were determined by straightforward umbrella
sampling. Statistics of the total density (including short-wavelength
fluctuations) within such probe volumes were then obtained using the
method outlined in Ref.\@~\cite{Varilly2011}.  Simulations with
spherical solutes required numerical calculation of the overlap
between cubic lattice cells and the solute's excluded volume.  These
overlap volumes were computed either with Monte Carlo integration or
with analytical expressions detailed in the SI. Though complicated,
these expressions significantly reduce computation time for solutes
that move continuously in space.

\bibliographystyle{pnas}
\bibliography{refs-lcw.bib}
\end{article}

\section{Supporting Information: Rough interfaces, accurate predictions: The necessity of
  capillary modes in a minimal model of nanoscale hydrophobic
  solvation}

\section{Integrating out Gaussian fluctuations in the presence of multiple ideal hydrophobes}

The presence of a solvent-excluding volume $v$ introduces a constraint on the
total solvent density field,
\begin{equation}\label{eq:constraint}
  \rho_l n(\mathbf{r}) + \delta\rho(\mathbf{r}) = 0,
\end{equation}
at each point $\mathbf{r}\in v$. Integrating out small-wavelength Gaussian
fluctuations $\delta\rho(\mathbf{r})$ in the presence of this constraint
requires evaluating
\begin{equation}\label{eq:integral}
  Z_v = \int\mathcal{D}\delta\rho(\mathbf{r})e^{-\beta H_s}\prod_{\mathbf{r}\in v}\delta\left(\rho_l n(\mathbf{r})+\delta\rho(\mathbf{r})\right).
\end{equation}
The product of delta functions enforces the constraint in Eq.\ \eqref{eq:constraint},
and $H_s$ is the unconstrained Hamiltonian for $\delta\rho(\mathbf{r})$,
\[
\beta H_s = \frac{1}{2}\int_{\mathbf{r}}\int_{\mathbf{r}'}\delta\rho(\mathbf{r})\chi^{-1}(\mathbf{r},\mathbf{r}')\delta\rho(\mathbf{r}').
\]
A formal expression for $Z_v$ can be easily derived, but numerical evaluation is
impractical due to the infinite product of delta functions. For this reason,
prior work on coarse graining water has replaced the pointwise constraint in
Eq.\ \eqref{eq:integral} by a single constraint on the average density in $v$, as in
Eq.\ [MT-4] of the main text. The resulting expression for $Z_v$ is much more manageable
and has been applied with great success to systems containing only a single solute
with a simple compact geometry. However, for systems with many independently moving
solutes that may be separated by large distances, the average constraint in
Eq.\ [MT-4] introduces spurious correlations between solutes that never vanish. If
$v$ is the union of $m$ non-intersecting regions,
\[
v = v^{(1)}\cup v^{(2)}\cup\dots\cup v^{(m)},
\]
a simple solution is to constrain the average density in each region
$v^{(\alpha)}$ separately, i.e., to apply the constraint expressed in
Eq.\ [MT-4] separately to each solute's excluded volume. The infinite product of
delta functions in Eq.\ \eqref{eq:integral} then becomes a product of $m$ delta
functions,
\[
\prod_{\mathbf{r}\in v}\delta\left(\rho_l n(\mathbf{r}) + \delta\rho(\mathbf{r})\right) \rightarrow \prod_{\alpha=1}^m \delta\left(\int_{\mathbf{r}\in v^{(\alpha)}}\rho_l n(\mathbf{r}) + \delta\rho(\mathbf{r})\right).
\]
Using the Fourier representation of the delta function, the integral in
Eq.\ \eqref{eq:integral} becomes
\[
\int\mathcal{D}\delta\rho(\mathbf{r})e^{-\beta H_s}\prod_{\alpha=1}^m\int \frac{d\psi_\alpha}{2\pi}\,\exp\left(-i\psi_\alpha\int_{\mathbf{r}\in v^{(\alpha)}}\rho_l n(\mathbf{r}) + \delta\rho(\mathbf{r})\right).
\]
Defining
\[
N_\alpha = \int_{\mathbf{r}\in v^{(\alpha)}}\rho_l n(\mathbf{r}) \quad\text{and}\quad \Phi(\mathbf{r}) = \begin{cases} i\psi_\alpha, & \mathbf{r}\in v^{(\alpha)}, \\ 0, & \text{else}, \end{cases}
\]
and rearranging the order of the integrals, we obtain
\[
\int\left(\prod_{\alpha=1}^m\frac{d\psi_\alpha}{2\pi}\,e^{-i N_\alpha\psi_\alpha}\right)\int\mathcal{D}\delta\rho(\mathbf{r})\exp\left(-\beta H_s-\int_\mathbf{r}\Phi(\mathbf{r})\delta\rho(\mathbf{r})\right).
\]
Evaluating the inner integral over $\delta\rho(\mathbf{r})$ results in
\[
Z_v = Z_0\int\left(\prod_{\alpha=1}^m \frac{d\psi_\alpha}{2\pi}\,e^{-i N_\alpha\psi_\alpha}\right)\exp\left(\frac{1}{2}\int_{\mathbf{r}}\int_{\mathbf{r}'}\Phi(\mathbf{r})\chi(\mathbf{r},\mathbf{r}')\Phi(\mathbf{r}')\right).
\]
The argument of the rightmost exponent evaluates to
\[
-\frac{1}{2}(2\pi)^2\sum_{\alpha=1}^m\sum_{\beta=1}^m \psi_\alpha(\chi_{\rm in})_{\alpha,\beta}\psi_\beta,
\]
where $(\chi_{\rm in})_{\alpha,\beta}$ is given by the double integral
\[
(\chi_{\rm in})_{\alpha,\beta} = \int_{\mathbf{r}\in v^{(\alpha)}}\int_{\mathbf{r}'\in v^{(\beta)}}\Theta(\mathbf{r})\chi(\mathbf{r},\mathbf{r}')\Theta(\mathbf{r}').
\]
With this, $Z_v$ can be expressed compactly in matrix notation,
\[
Z_v = Z_0\int\left(\prod_{\alpha=1}^m\frac{d\psi_\alpha}{2\pi}\right)\exp\left(-\frac{1}{2}\overline{\psi}^\top\chi_{\rm in}\overline{\psi}-i\overline{N}^\top\overline{\psi}\right),
\]
where $\overline{\psi}$ is a column vector with elements $\psi_\alpha$,
$\overline{N}$ is a column vector with elements $N_\alpha$, and
$\chi_{\rm in}$ is an $m\times m$ matrix with elements
$(\chi_{\rm in})_{\alpha,\beta}$. This integral is easily evaluated to give
\[
Z_v = Z_0\frac{1}{\sqrt{\det(2\pi\chi_{\rm in})}}\exp\left(-\frac{1}{2}\overline{N}^\top\chi^{-1}_{\rm in}\overline{N}\right).
\]
The free energetic contribution of the small-wavelength field, relative to the
unconstrained case, is then given by
\begin{equation}\label{eq:mbasisFE}
  F_v - F_0 = -T\ln\left(\frac{Z_v}{Z_0}\right) = \frac{T}{2}\left(\ln\left(\det(2\pi\chi_{\rm in})\right) + \overline{N}^\top\chi_{\rm in}^{-1}\overline{N}\right),
\end{equation}
which are the final two terms in Eq.\ [MT-5] of the main text. For two solutes
$\alpha$ and $\beta$, the off-diagonal matrix element
$(\chi_{\rm in})_{\alpha,\beta}$ gives the coupling between Gaussian
fluctuations in regions $v^{(\alpha)}$ and $v^{(\beta)}$. If the solutes are
separated by much more than the correlation length of water, this term vanishes.
In the limit of infinite separation between all solutes, $\chi_{\rm in}$ becomes a
diagonal matrix and the free energy in Eq.\ \eqref{eq:mbasisFE} reduces to a simple
sum of uncoupled terms,
\[
F_v - F_0 = \frac{T}{2}\sum_{\alpha=1}^m\left(\ln(2\pi(\chi_{\rm in})_{\alpha,\alpha}) + \frac{N_\alpha^2}{(\chi_{\rm in})_{\alpha,\alpha}}\right).
\]

\section{Exact expression for the overlap between the lattice and spherical solute}

The Hamiltonian given in Eq.\ [MT-5] of the main text requires that we compute $v_i,$ the overlap
between the solute's excluded volume and the cubic lattice cells.
The overlap can be estimated accurately using Monte Carlo integration when the solute
does not move over the course of the simulation. However, it is also possible to
compute the overlap analytically, a pragmatic approach when the solute is mobile.
This strategy improves accuracy and saves significant 
computation.

In order to efficiently compute the overlap between cells of the lattice and spherical
objects, we derived an exact formula using ordinary calculus. There are eight distinct
cases to consider, each with several sub-cases. The calculation is a tedious, but
straightforward exercise in enumerating the cases. There are two main, equivalent
scenarios: first, if the cubic region lies entirely within the solute, then the 
overlap is simply $\delta^3.$ Otherwise, the diagonal of the cubic cell, which we 
refer to as the line segment connecting $(x_l,y_l,z_l)$ to $(x_u,y_u,z_u)$ throughout,
does not lie entirely within the sphere. Every sub-case can be rotated and decomposed
into one of these two sub-cases. As such, knowledge of the following specific case 
is sufficient to compute the overlap in general.

Consider a sphere of radius $R$ centered at the origin. We will work through the case
that $R>\delta$, where $\delta$ is the coarse-graining length (the side length of the cubes
that form the lattice). The overlap between the sphere and a cubic lattice cell 
can always be decomposed into contributions from each of the eight octants 
of the sphere. Without loss of generality, we will select the first octant $x,y,z > 0$, 
and assume that the diagonal of the cubic lattice cell lies entirely within the octant.

With this particular case fully specified, we now carry out the integral

\begin{equation}
  \int_0^{\sqrt{R^2-z^2-y^2}} \int_0^{\sqrt{R^2-z^2}} \int_0^R dx\ dy\ dz\ 
  -\int_{x_l}^{\sqrt{R^2-z^2-y^2}} \int_{y_l}^{\sqrt{R^2-z^2}} \int_{z_l}^R dx\ dy\ dz.
  \label{eq:overlap}
\end{equation}

This expression is valid when $x_u,y_u,z_u > R$. If this is not the case, then additional
integrals that account for the volume of the sphere in the first octant that lies 
outside the cubic cell must be subtracted away. These additional integrals have the same
form as the integral in Eq.\ \eqref{eq:overlap}.

The integral specifying the overlap can be calculated analytically. We have not been 
able to simplify the expression into a convenient form, but the analytical result can
easily be used within computer code.

The simplified expression for Eq.\ \eqref{eq:overlap} is
\begin{align}
  & \frac{1}{2} \left[ -\frac{1}{3} 2 \tan ^{-1}\left(\frac{x_l y_l}{R \sqrt{R^2-x_l^2-y_l^2}}\right) R^3-2 x_l y_l z_l-\frac{1}{3} x_l \left(x_l^2-3 R^2\right) \tan ^{-1}\left(\frac{y_l}{\sqrt{R^2-x_l^2-y_l^2}}\right) \right.\\
  \nonumber & -\frac{1}{2} y_l \left(R^2-y_l^2\right) \tan^{-1}\left(\frac{x_l \sqrt{R^2-x_l^2-y_l^2}}{-R^2+x_l^2+y_l^2}\right)-\frac{1}{6} y_l \left(3 R^2+y_l^2\right) \tan^{-1}\left(\frac{x_l \sqrt{R^2-x_l^2-y_l^2}}{-R^2+x_l^2+y_l^2}\right)\\
\nonumber &  +\frac{1}{3} x_l \left(x_l^2-3 R^2\right) \tan^{-1}\left(\frac{\sqrt{R^2-x_l^2-z_l^2}}{\sqrt{z_l^2}}\right) - z_l \left(R^2-z_l^2\right) \tan^{-1}\left(\frac{x_l \sqrt{R^2-x_l^2-z_l^2}}{-R^2+x_l^2+z_l^2}\right)\\
\nonumber &  +\frac{1}{2} \sqrt{z_l^2} \left(R^2-z_l^2\right) \tan^{-1}\left(\frac{x_l \sqrt{R^2-x_l^2-z_l^2}}{-R^2+x_l^2+z_l^2}\right)\\
\nonumber & \left. +\frac{2}{3} x_l y_l \sqrt{R^2-x_l^2-y_l^2}+x_l z_l \sqrt{R^2-x_l^2-z_l^2}-\frac{1}{3} x_l \sqrt{z_l^2} \sqrt{R^2-x_l^2-z_l^2}\right] \\
  \nonumber &  +\frac{1}{2} \left[\frac{2}{3} \tan ^{-1}\left(\frac{x_u y_l}{R \sqrt{R^2-x_u^2-y_l^2}}\right) R^3+2 x_u y_l z_l+\frac{1}{3} x_u \left(x_u^2-3 R^2\right) \tan^{-1}\left(\frac{y_l}{\sqrt{R^2-x_u^2-y_l^2}}\right) \right.\\
\nonumber &  +\frac{1}{2} y_l \left(R^2-y_l^2\right) \tan^{-1}\left(\frac{x_u \sqrt{R^2-x_u^2-y_l^2}}{-R^2+x_u^2+y_l^2}\right)+\frac{1}{6} y_l \left(3 R^2+y_l^2\right) \tan^{-1}\left(\frac{x_u \sqrt{R^2-x_u^2-y_l^2}}{-R^2+x_u^2+y_l^2}\right) \\
\nonumber & -\frac{1}{3} x_u \left(x_u^2-3 R^2\right) \tan^{-1}\left(\frac{\sqrt{R^2-x_u^2-z_l^2}}{\sqrt{z_l^2}}\right)+z_l \left(R^2-z_l^2\right) \tan^{-1}\left(\frac{x_u \sqrt{R^2-x_u^2-z_l^2}}{-R^2+x_u^2+z_l^2}\right) \\
\nonumber & \left. -\frac{1}{2} \sqrt{z_l^2} \left(R^2-z_l^2\right) \tan^{-1}\left(\frac{x_u \sqrt{R^2-x_u^2-z_l^2}}{-R^2+x_u^2+z_l^2}\right)-\frac{2}{3} x_u y_l \sqrt{R^2-x_u^2-y_l^2}-x_u z_l \sqrt{R^2-x_u^2-z_l^2}+\frac{1}{3} x_u \sqrt{z_l^2} \sqrt{R^2-x_u^2-z_l^2}\right] \\
  \nonumber & +\frac{1}{2} \left[ -\frac{2}{3} \tan^{-1}\left(\frac{\frac{x_l z_l}{R \sqrt{R^2-x_l^2-z_l^2}}-\frac{x_u z_l}{R \sqrt{R^2-x_u^2-z_l^2}}}{\frac{x_l x_u z_l^2}{R^2 \sqrt{R^2-x_l^2-z_l^2} \sqrt{R^2-x_u^2-z_l^2}}+1}\right) R^3-\frac{1}{6} \left(z_l \left(3 R^2+z_l^2\right)\right) \tan^{-1}\left(2 z_l \sqrt{R^2-x_l^2-z_l^2},-2 x_l z_l\right) \right. \\
\nonumber & \left. +\frac{1}{6} \left(z_l \left(3 R^2+z_l^2\right)\right) \tan^{-1}\left(2 z_l \sqrt{R^2-x_u^2-z_l^2},-2 x_u z_l\right)\right].
\end{align}

\begin{figure*}[tbp]
\begin{center}
\includegraphics[width=\textwidth]{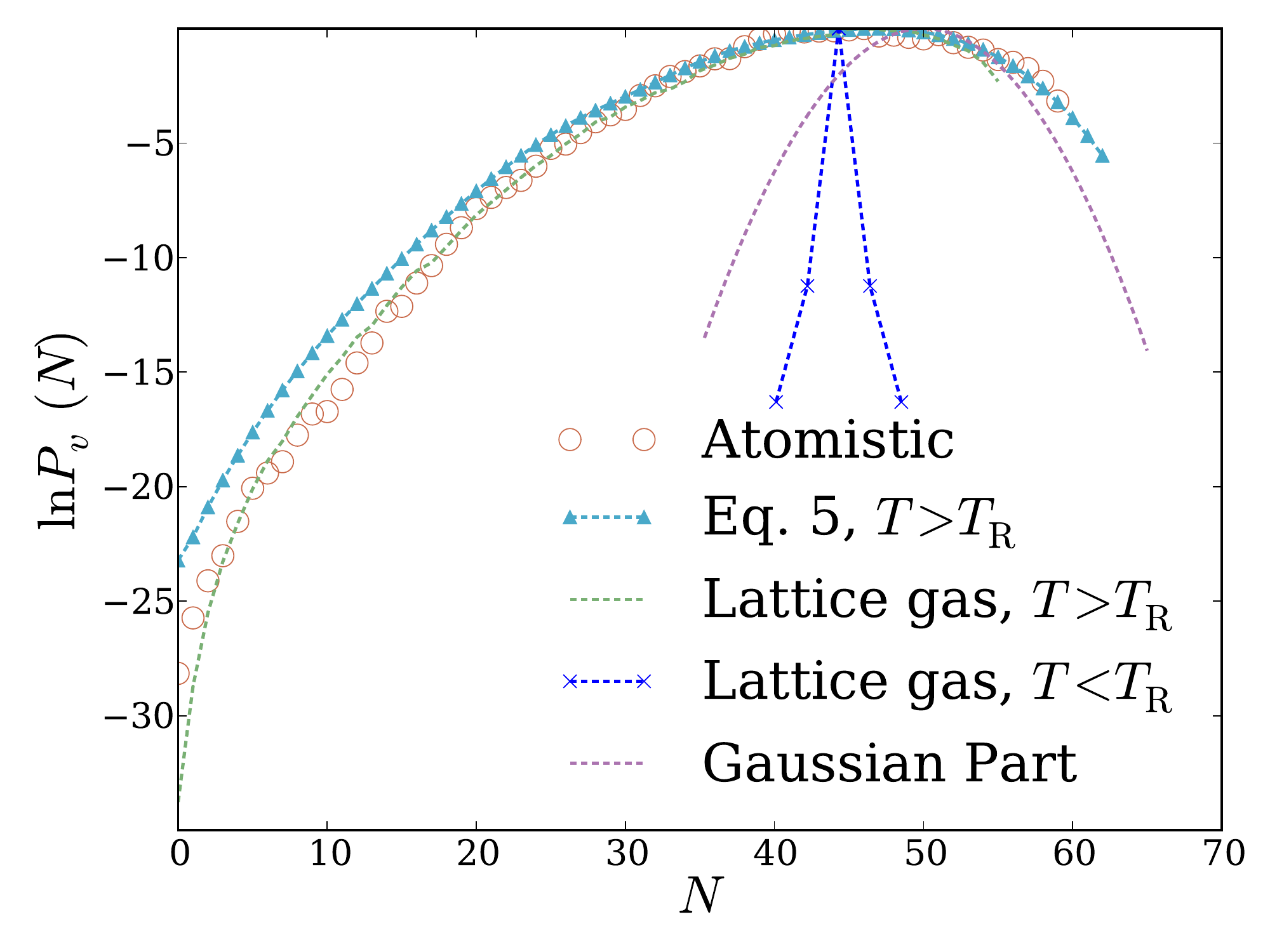}
\end{center}
\caption{ 
  The probability of finding $N$ molecules in a probe volume situated
  at a liquid-vapor interface. Here, we show the Gaussian part of the Hamiltonian 
  given in the main text,
  Eq.\ [MT-5], plotted alongside results from atomistic simulations, 
  the coarse-grained model of Eq.\ [MT-5], a rough lattice gas, and a cold lattice gas at coexistence.
}
\label{fig:SIinterface}
\end{figure*}

\begin{figure*}[tbp]
\begin{center}
\includegraphics[width=\textwidth]{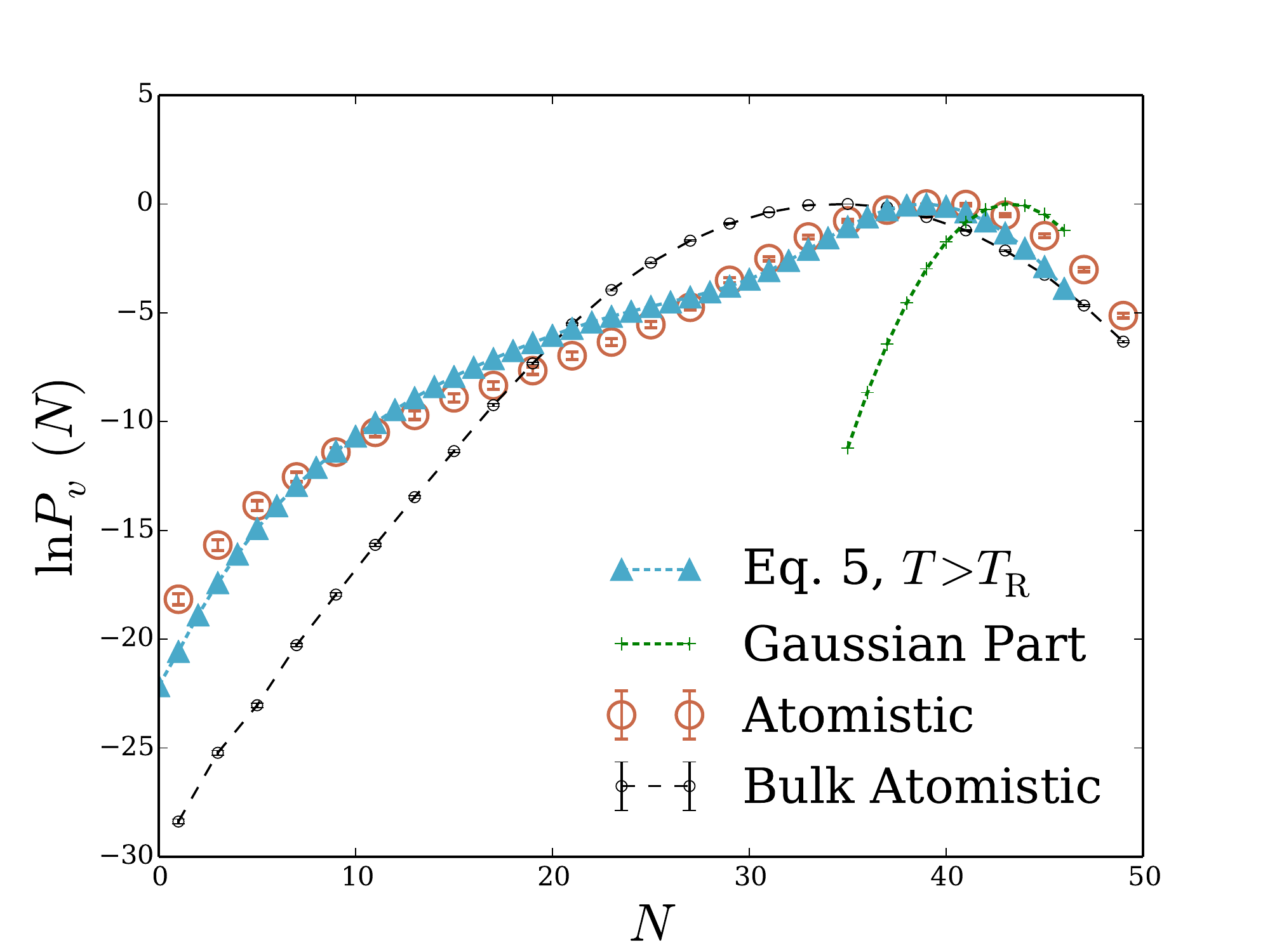}
\end{center}
\caption{ 
  The probability of finding $N$ molecules in a probe volume between two
  ideal, 20.24\AA\ $\times$ 5.52\AA\ $\times$ 9.2\AA\ hydrophobic plates,
  separated by a distance of 5.52\AA.
  Here, we show the Gaussian part of the Hamiltonian given in the main text,
  Eq.\ [MT-5], plotted alongside results from atomistic simulation, 
  the coarse-grained model of Eq.\ [MT-5], and the bulk $P_v(N)$ for a probe volume 
  of the same size as the region between the walls. Note that the Gaussian
  part of the Hamiltonian overcompensates for the shift towards higher
  density.
}
\label{fig:SIwall1}
\end{figure*}

\begin{figure*}[tbp]
\begin{center}
\includegraphics[width=\textwidth]{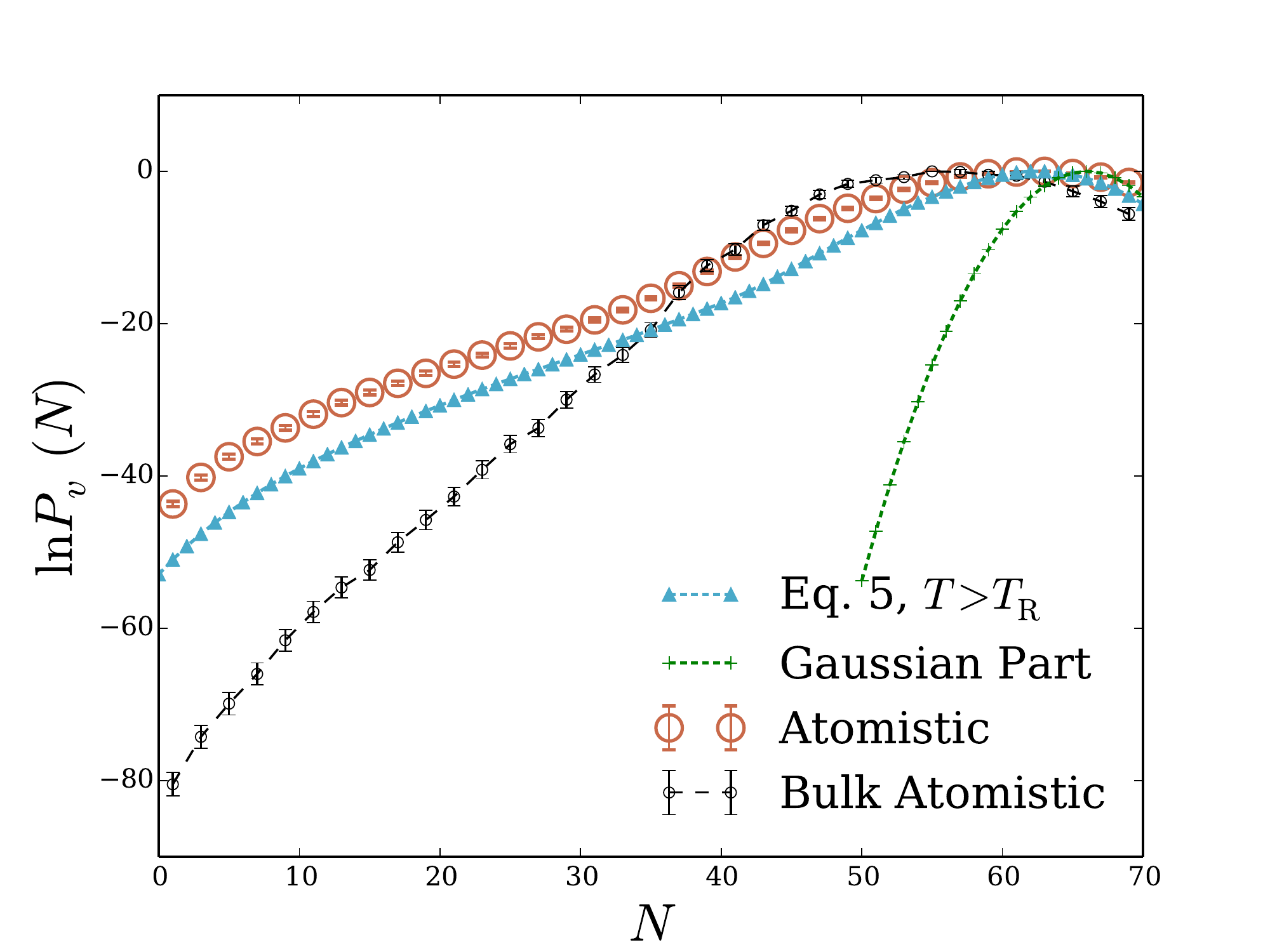}
\end{center}
\caption{ 
  The probability of finding $N$ molecules in a probe volume between two
  ideal, 20.24\AA\ $\times$ 5.52\AA\ $\times$ 9.2\AA\ hydrophobic plates,
  separated by a distance of 9.2\AA.
  Here, we show the Gaussian part of the Hamiltonian given in the main text,
  Eq.\ [MT-5], plotted alongside the results from atomistic simulation, 
  the coarse-grained model of Eq.\ [MT-5], and the bulk $P_v(N)$ for a probe volume 
  of the same size as the region between the walls. Note that the Gaussian
  part of the Hamiltonian overcompensates for the shift towards higher
  density.
}
\label{fig:SIwall2}
\end{figure*}

\end{document}